\documentclass[sigconf]{acmart}

\usepackage{balance}
\usepackage{amsmath}
\DeclareMathOperator*{\argmax}{arg\,max}

\copyrightyear{2019}
\acmYear{2019}
\setcopyright{rightsretained}
\acmConference[CIKM '19] {1st International Workshop on Graph Representation Learning and its Applications}{November 3rd--7th, 2019}{Beijing, China}
\acmDOI{}
\acmISBN{}

\usepackage{subfigure}
\usepackage{paralist}
\usepackage{textcomp}

\begin{document}

%
\title[NISER: Normalized Item and Session Representations with Graph Neural Networks]{NISER: Normalized Item and Session Representations to Handle Popularity Bias}

\author{ Priyanka Gupta, Diksha Garg, Pankaj Malhotra, Lovekesh Vig, Gautam Shroff}
\email{ priyanka.g35@tcs.com, diksha.7@tcs.com, malhotra.pankaj@tcs.com, lovekesh.vig@tcs.com, gautam.shroff@tcs.com}
\affiliation{%
  \institution{TCS Research, New Delhi, India}
}

\renewcommand{\shortauthors}{P. Gupta, Di. Garg, et al.}

%

\begin{abstract}
The goal of session-based recommendation (SR) models is to utilize the information from past actions (e.g. item/product clicks) in a session to recommend items that a user is likely to click next.
Recently it has been shown that the sequence of item interactions in a session can be modeled as graph-structured data to better account for  complex item transitions.
Graph neural networks (GNNs) can learn useful representations for such session-graphs, and have been shown to improve over sequential models such as recurrent neural networks \cite{wu2018session}.
However, we note that these GNN-based recommendation models suffer from popularity bias: the models are biased towards recommending popular items, and fail to recommend relevant long-tail items (less popular or less frequent items). 
Therefore, these models perform poorly for the less popular new items arriving daily in a practical online setting. We demonstrate that this issue is, in part, related to the magnitude or norm of the learned item and session-graph representations (embedding vectors). 
We propose a training procedure that mitigates this issue by using normalized representations.
The models using normalized item and session-graph representations perform significantly better: 
i. for the less popular long-tail items in the offline setting, and 
ii. for the less popular newly introduced items in the online setting. 
Furthermore, our approach significantly improves upon existing state-of-the-art on three benchmark datasets.
\end{abstract}

%
%
%

%
%
\maketitle

\section{Introduction}
\begin{figure}
	\centering
	\includegraphics[trim={0.2cm 0cm 0.4cm 0cm},clip,width=0.3\textwidth]{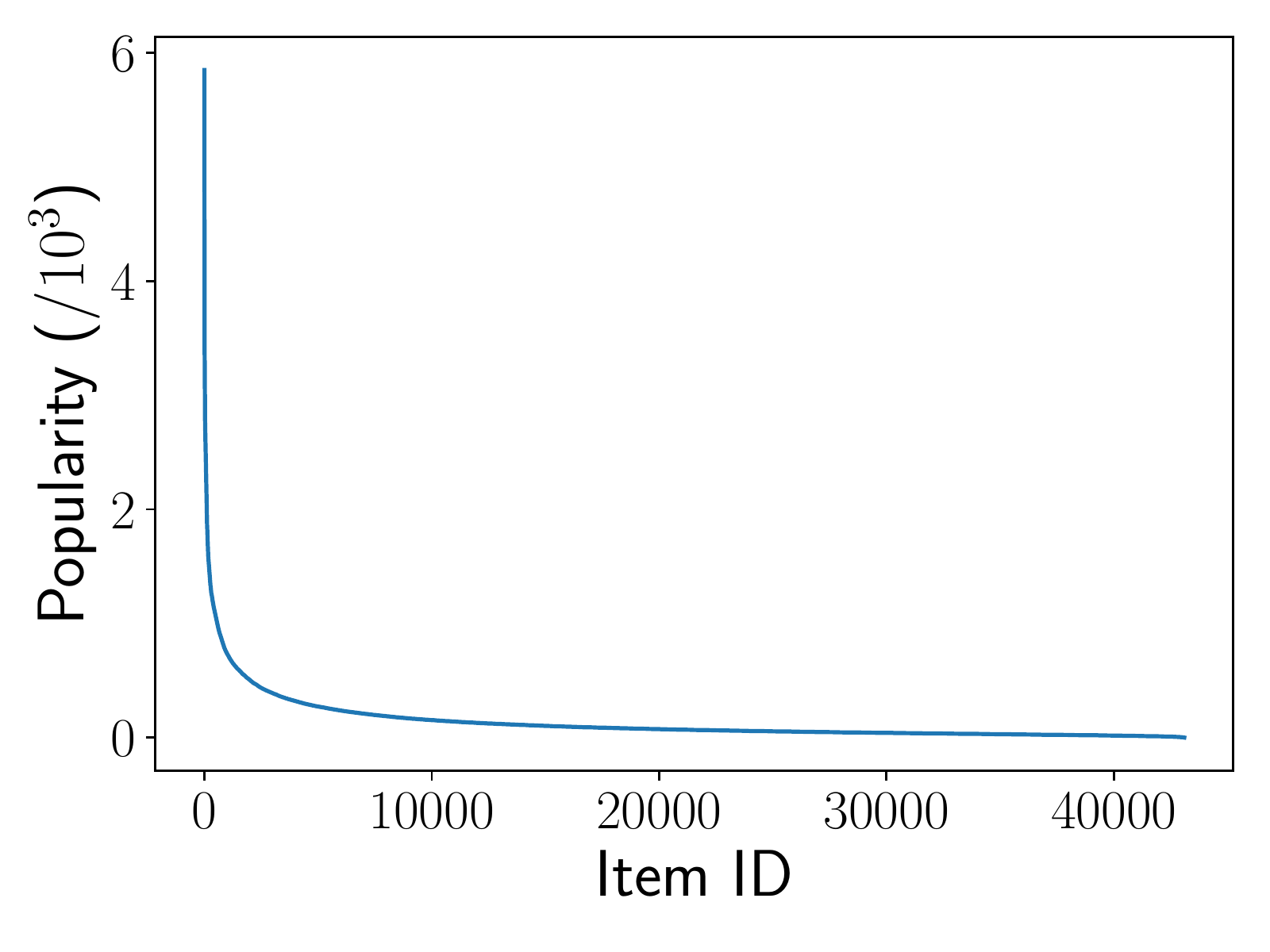}
	\vspace{-2mm}
	\caption{Typical popularity distribution of items depicting the long tail.\label{fig:freq_bias}}
	\vspace{-4mm}
\end{figure}

The goal of session-based recommendation (SR) models is to recommend top-K items to a user based on the sequence of items clicked so far.
Recently, several effective models for SR based on deep neural networks architectures have been proposed \cite{wu2018session,liu2018stamp,li2017neural,wang2019collaborative}.
These approaches consider SR as a multi-class classification problem where input is a sequence of items clicked in the past in a session and target classes correspond to available items in the catalog. 
Many of these approaches use sequential models like recurrent neural networks considering a session as a sequence of item click events \cite{hidasi2015session,jannach2017recurrent,wang2019collaborative}.
On the other hand, approaches like STAMP \cite{liu2018stamp} consider a session to be a set of items, and use attention models while learning to weigh (attend to) items as per their relevance to predict the next item.
Approaches like NARM \cite{li2017neural} and CSRM \cite{wang2019collaborative} use a combination of sequential and attention models.

An important building block in most of these deep learning approaches is their ability to learn representations or embeddings for items and sessions.
Recently, an alternative approach, namely SR-GNN \cite{wu2018session}, has been proposed to model the sessions as graph-structured data using GNNs \cite{li2015gated} rather than as sequences or sets, noting that users tend to make complex \textit{to-and-fro} transitions across items within a session:
for example, consider a session $s = i_1,i_2,i_1,i_3,i_4$ of item clicks by a user. Here, the user clicks on item $i_1$, then clicks on item $i_2$ and then re-clicks on item $i_1$. 
This sequence of clicks induces a graph where nodes and edges correspond to items and transitions across items, respectively.
For session $s$ in the above example, the fact that $i_2$ and $i_3$ are neighbors of $i_1$ in the induced session-graph, the representation of $i_1$ can be updated using representations of its neighbors, i.e. $i_2$ and $i_3$, and thus obtain a more context-aware and informative representations. 
It is worth noting that this way of capturing neighborhood information has also been found to be effective in neighborhood-based SR methods such as SKNN \cite{jannach2017recurrent} and STAN \cite{garg2019sequence}.

It is well-known that more popular items are presented and interacted-with more often on online platforms. This results in a skewed distribution of items clicked by users  \cite{steck2011item,abdollahpouri2017controlling,yang2018unbiased}, as illustrated in Fig. \ref{fig:freq_bias}.
The models trained using the resulting data tend to have \textit{popularity bias}, i.e. they tend to recommend more popular items over rarely clicked items. 

We note that SR-GNN (referred to as GNN hereafter) also suffers from popularity bias. 
This problem is even more severe in a practical online setting where new items are frequently added to the catalog, and are inherently less popular during initial days.
To mitigate this problem, we study GNN through the lens of an \textit{item and session-graph representation learning mechanism}, where the goal is to obtain a session-graph representation that is \textit{similar} to the representation of the item likely to be clicked next.
We motivate the advantage of restricting the item and session-graph representations to lie on a unit hypersphere both during training and inference, and propose \textbf{NISER}: \textbf{N}ormalized \textbf{I}tem and \textbf{Se}ssion \textbf{R}epresentations model for SR. 
We demonstrate the enhanced ability of NISER to deal with popularity bias in comparison to a vanilla GNN model in the offline as well as online settings. 
We also extend NISER to incorporate the sequential nature of a session via position embeddings \cite{vaswani2017attention}, thereby leveraging the benefits of both sequence-aware models (like RNNs) and graph-aware models.

\section{Related Work}

Recent results in computer vision literature, e.g. \cite{wang2017normface,zheng2018ring}, indicate the effectiveness of normalizing the final image features during training, and argue in favor of cosine similarity over inner product for learning and comparing feature vectors. 
\cite{zheng2018ring} introduces the ring loss for soft feature normalization which eventually learns to constrain the feature vectors on a unit hypersphere. Normalizing words embeddings is also popular in NLP applications, e.g. \cite{peng2015comparative} proposes penalizing the L$_2$ norm of word embeddings for regularization. 
However, to the best of our knowledge, the idea of normalizing item and session-graph embeddings or representations has not been explored. 

In this work, we study the effect of normalizing the embeddings on popularity bias which has not been established and studied so far.
Several approaches to deal with popularity bias exist in the collaborative filtering settings, e.g. \cite{abdollahpouri2017controlling,steck2011item,yang2018unbiased}.
To deal with popularity bias, \cite{abdollahpouri2017controlling} introduces the notion of flexible regularization in a learning-to-rank algorithm. Similarly \cite{steck2011item,yang2018unbiased} uses the power-law of popularity where the probability of recommending an item is a smooth function of the items' popularity, controlled by an exponent factor. However, to the best of our knowledge, this is the first attempt to study and address popularity bias in DNN-based SR using SR-GNN \cite{wu2018session} as a working example.  
Furthermore, SR-GNN does not incorporate the sequential information explicitly to obtain the session-graph representation. We study the effect of incorporating position embeddings \cite{vaswani2017attention} and show that it leads to minor but consistent improvement in recommendation performance.

\section{Problem Definition\label{sec:prob_def}}		
Let $\mathcal{S}$ denote all past sessions, and $\mathcal{I}$ denote the set of $m$ items observed in the set $\mathcal{S}$.
Any session $s \in \mathcal{S}$ is a chronologically ordered tuple of item-click events: 
$s = (i_{s,1},i_{s,2},\ldots,i_{s,l})$, where each of the $l$ item-click events $i_{s,j}$ ($j=1\ldots l$) corresponds to an item in $\mathcal{I}$, and $j$ denotes the position of the item $i_{s,j}$ in the session $s$.
A session $s$ can be modeled as a graph $\mathcal{G}_s = (\mathcal{V}_s, \mathcal{E}_s)$, where $i_{s,j}\in \mathcal{V}_s$ is a node in the graph. 
Further, $(i_{s,j},i_{s,j+1})\in \mathcal{E}_s$ is a directed edge from $i_{s,j}$ to $i_{s,j+1}$.
Given $s$, the goal of SR is to predict the next item $i_{s,l+1}$ by estimating the $m$-dimensional probability vector $\mathbf{\hat{y}}_{s,l+1}$ corresponding to the relevance scores for the $m$ items. 
The $K$ items with highest scores constitute the top-K recommendation list.

\section{Learning Item and Session Representations\label{sec:approach}}
Each item is mapped to a $d$-dimensional vector from the trainable embedding look-up table or matrix $\mathbf{I} = [\mathbf{i}_1,\mathbf{i}_2,\ldots,\mathbf{i}_m]^T \in \mathbb{R}^{m\times d}$ such that each row $\mathbf{i}_j\in \mathbb{R}^d$ is the $d$-dimensional representation or embedding\footnote{We use the terms representation and embedding interchangeably.} vector corresponding to item $i_j \in \mathcal{I}$ ($j=1\ldots m$). 
Consider any function $f$ (e.g. a neural network as in \cite{liu2018stamp,wu2018session})---parameterized by $\boldsymbol\theta$---that maps the items in a session $s$ to session embedding $\mathbf{s} = f(\mathbf{I}_s;\boldsymbol\theta)$, where\footnote{To ensure same dimensions of $\mathbf{I}_s \in \mathbb{R}^{L\times d}$ across sessions, we can pad with a dummy vector $L-l$ times.} $\mathbf{I}_s = [\mathbf{i}_{s,1},\mathbf{i}_{s,2},\ldots,\mathbf{i}_{s,l}]^T \in \mathbb{R}^{l\times d}$.
Along with $\mathbf{I}_s$ as an input which considers $s$ as a sequence, we also introduce an adjacency matrix $\mathbf{A}_s$ to incorporate the graph structure. We discuss this in more detail later in Section \ref{ssec:niser-gnn}.

The goal is to obtain $\mathbf{s}$ that is close to the embedding $\mathbf{i}_{s,l+1}$ of the target item $i_{k}=i_{s,l+1}$, such that the estimated index/class for the target item is $k = \argmax_j~ \mathbf{i}_j^T\mathbf{s}$ with $j=1\ldots m$. 
In a DNN-based model $f$, this is approximated via a differentiable softmax function such that the probability of next item being $i_k$ is given by:
\begin{equation}
\label{eq:softmax1}
p_k(\mathbf{s}) = \hat{\mathbf{y}}_k = \frac{\text{exp}(\mathbf{i}_k^T\mathbf{s})}{\sum_{j=1}^{m}\text{exp}(\mathbf{i}_j^T\mathbf{s})}.
\end{equation}

For this $m$-way classification task, softmax (cross-entropy) loss is used during training for estimating $\boldsymbol \theta$ by minimizing the sum of $\mathcal{L(\hat{\mathbf{y}})} = - \sum_{j=1}^{m}\mathbf{y}_j\text{log}(\hat{\mathbf{y}}_j)$
over all training samples, where $\mathbf{y} \in \{0,1\}^m$ is a 1-hot vector with  $\mathbf{y}_k=1$ corresponding to the correct (target) class $k$. 

We next introduce the radial property \cite{wang2017normface,zheng2018ring} of softmax loss, and then use it to motivate the need for normalizing the item and session representations in order to reduce popularity bias.

\subsection{Radial Property of Softmax Loss}

\begin{figure*}[h]
	\includegraphics[scale=0.5,trim={1cm 4.5cm 2cm 4cm},clip]{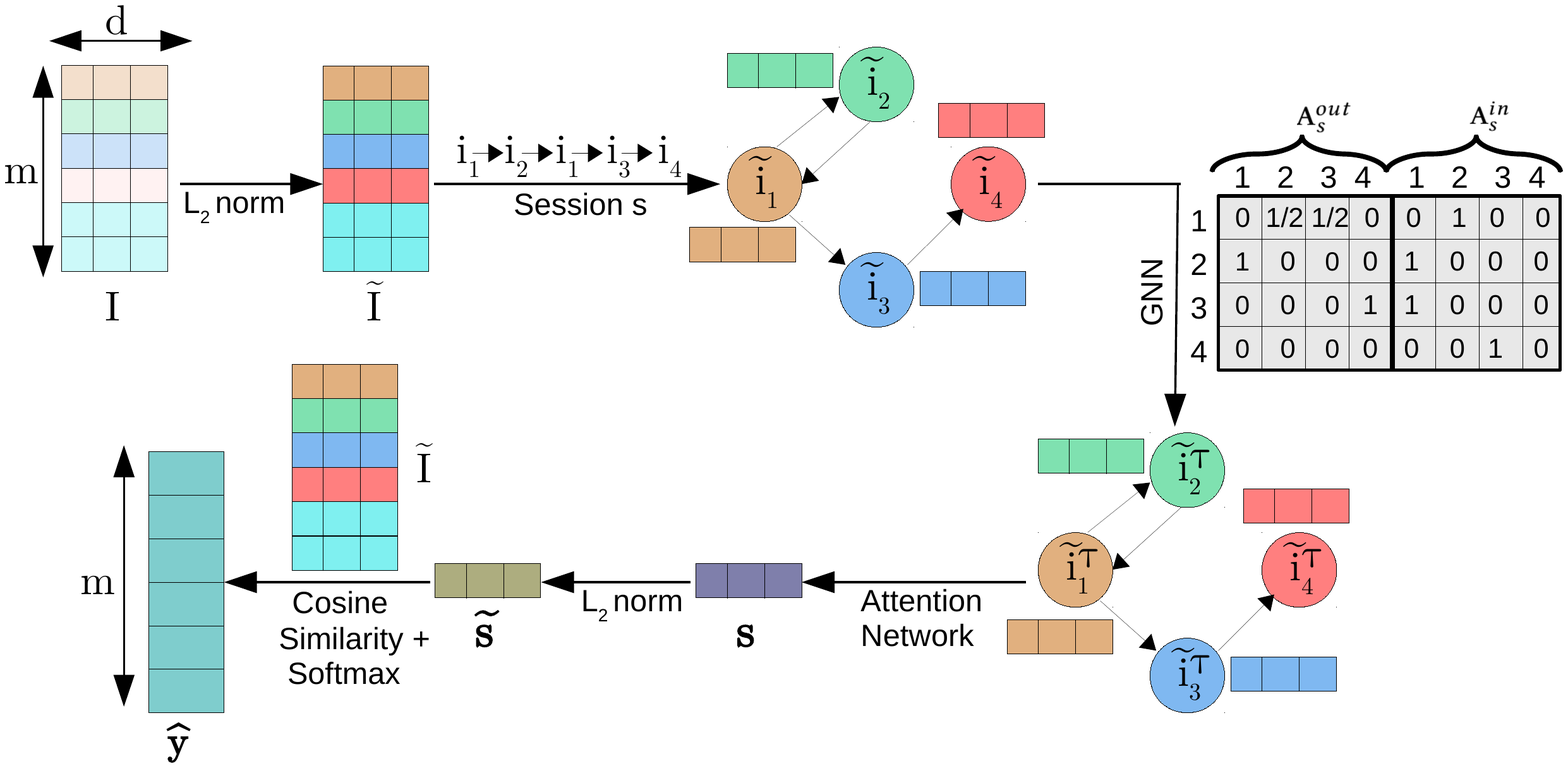}
	\caption{Illustrative flow diagram for NISER.\label{fig:niser}}
\end{figure*}

It is well-known that optimizing for the softmax loss leads to a radial distribution of features for a target class \cite{wang2017normface,zheng2018ring}: 
If $k = \argmax_j~ \mathbf{i}_j^T\mathbf{s}$, then it is easy to show that 
\begin{equation}
\frac{\text{exp}(\sigma \mathbf{i}_k^T\mathbf{s})}{\sum_{j\neq k}\text{exp}(\mathbf{i}_j^T\mathbf{s})~+~\text{exp}(\sigma \mathbf{i}_k^T\mathbf{s})} > \frac{\text{exp}(\mathbf{i}_k^T\mathbf{s})}{\sum_{j=1}^{m}\text{exp}(\mathbf{i}_j^T\mathbf{s})}
\end{equation}
for any $\sigma>1$.
This, in turn, implies that softmax loss favors large norm of features for easily classifiable instances.
We omit details for brevity and refer the reader to \cite{wang2017normface,zheng2018ring} for a thorough analysis and proof.
This means that a high value of $\hat{\mathbf{y}}_k$ can be attained by multiplying vector $\mathbf{i}_k$ by a scalar value $\sigma >1$; or simply by ensuring a large norm for the item embedding vector.
\begin{figure}[h]
	\centering
	\includegraphics[trim={0.25cm 0cm 0.4cm 0cm},clip,width=0.35\textwidth]{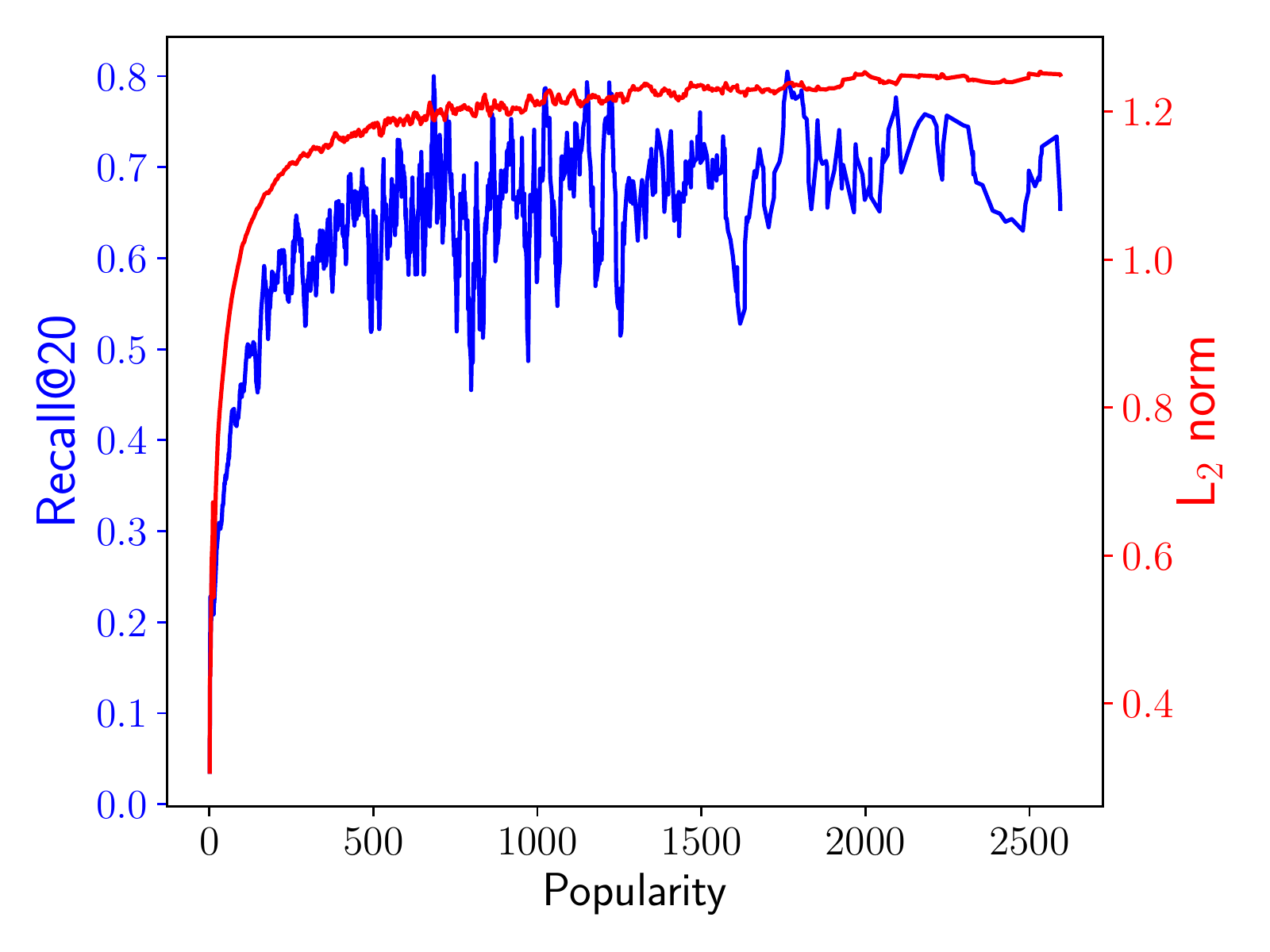}
	\vspace{-2mm}
	\caption{Recall@20 and L$_2$ norm of learned item embeddings decreases with decreasing popularity in GNN \cite{wu2018session}.\label{fig:pop}}
	\vspace{-4mm}
\end{figure}
\subsection{Normalizing the Representations \label{ssec:motiv}}
We note that the radial property has an interesting implication in the context of popularity bias: target items that are easier to predict are likely to have higher L$_2$ norm. 
We illustrate this with the help of an example: Items that are popular are likely to be clicked more often, and hence the trained parameters $\boldsymbol{\theta}$ and $\mathbf{I}$ should have values that ensure these items get recommended more often.
The radial property implies that for a given input $s$ and a popular target item $i_k$, a correct classification decision can be obtained as follows: learn the embedding vector $\mathbf{i}_k$ with high $||\mathbf{i}_k||_2$ such that the inner product $\mathbf{i}_k^T\mathbf{s} = ||\mathbf{i}_k||_2||\mathbf{s}||_2cos\alpha$ (where $\alpha$ is the angle between the item and session embeddings) is likely to be high to ensure large value for $\hat{\mathbf{y}}_k$ (even when $\alpha$ is not small enough and $||s||_2$ is not large enough). 
When analyzing the item embeddings from a GNN model \cite{wu2018session}, we observe that this is indeed the case: As shown in Fig. \ref{fig:pop}, items with high popularity have high L$_2$ norm while less popular items have significantly lower L$_2$ norm. 
Further, performance of GNN (depicted in terms of Recall@20) degrades as the popularity of the target item decreases.

\subsection{NISER}
Based on the above observations, we consider to minimize the influence of embedding norms in the final classification and recommendation decision. We propose optimizing for cosine similarity as a measure of similarity between item and session embeddings instead of the above-stated inner product. 
Therefore, during training as well as inference, we normalize the item embeddings as 
$\tilde{\mathbf{i}}_k = \frac{\mathbf{i}_{k}}{||\mathbf{i}_{k}||_{_2}}$, 
and use them to get $\tilde{\mathbf{I}}_s$.
The session embedding is then obtained as 
$\mathbf{s} = f(\tilde{\mathbf{I}}_s;\boldsymbol\theta)$, and is similarly normalized to $\tilde{\mathbf{s}}$  to enforce a unit norm.
The normalized item and session embeddings are then used to obtain the relevance score for next clicked item $i_k$ computed as, 
\begin{equation} \label{eq:softmax}
\hat{\mathbf{y}}_k=\frac{\text{exp}(\sigma\tilde{\mathbf{i}}_k^T\tilde{\mathbf{s}})}{\sum_{j=1}^{m}\text{exp}(\sigma \tilde{\mathbf{i}}_j^T\tilde{\mathbf{s}})}.
\end{equation}
Note that the cosine similarity $\tilde{\mathbf{i}}_k^T\tilde{\mathbf{s}}$ is restricted to $[-1,1]$. As shown in \cite{wang2017normface}, this implies that the softmax loss is likely to get saturated at high values for the training set: a scaling factor $\sigma>1$ is useful in practice to allow for better convergence.
\section{Leveraging NISER with GNN\label{ssec:niser-gnn}}
We consider learning the representations of items and session-graphs with GNNs where the session-graph is represented by $\mathcal{G}_s$ as introduced in Section \ref{sec:prob_def}. 
Consider two normalized adjacency matrices $\mathbf{A}^{in}_s \in \mathbb{R}^{l,l}$ and $\mathbf{A}^{out}_s \in \mathbb{R}^{l,l}$ corresponding to the incoming and outgoing edges in graph $\mathcal{G}_s$ as illustrated in Fig. \ref{fig:niser}.
GNN takes adjacency matrices $\mathbf{A}^{in}_s$ and $\mathbf{A}^{out}_s$, and the normalized item embeddings $\tilde{\mathbf{I}}_s$ as input, and returns an updated set of embeddings after $\tau$ iterations of message propagation across vertices in the graph using gated recurrent units \cite{li2015gated}: $[\tilde{\mathbf{i}}^\tau_{s,1},\tilde{\mathbf{i}}^\tau_{s,2},\ldots,\tilde{\mathbf{i}}^\tau_{s,l}] = G(\mathbf{A}^{in}_s,\mathbf{A}^{out}_s,\tilde{\mathbf{I}}_s;\boldsymbol{\theta}_g)$, where $\boldsymbol{\theta}_g$ represents the parameters of the GNN function $G$.
For any node in the graph, the current representation of the node and the representations of its neighboring nodes are used to iteratively update the representation of the node $\tau$ times.
More specifically, the representation of node $i_{s,j}$ in the $t$-th message propagation step is updated as follows:


\begin{align}
\label{eq:adj}
\mathcal{\mathbf{a}}_{s,j}^t &= [\mathcal{\mathbf{A}}^{in}_{s,j:} \tilde{\mathbf{I}}^{t-1}_s \mathbf{H}_1, \mathcal{\mathbf{A}}^{out}_{s,j:} \tilde{\mathbf{I}}^{t-1}_s \mathbf{H}_2]^T + \mathbf{b},\\
\mathcal{\mathbf{z}}_{s,j}^t &= \sigma (\mathbf{W}_z \mathcal{\mathbf{a}}_{s,j}^t + \mathbf{U}_z\tilde{\mathbf{i}}_{s,j}^{t-1}),\\
\mathbf{r}_{s,j}^t &= \sigma (\mathbf{W}_r \mathcal{\mathbf{a}}_{s,j}^t + \mathbf{U}_r\tilde{\mathbf{i}}_{s,j}^{t-1}),\\
\hat{\mathbf{i}}_{s,j}^t &= \tanh (\mathbf{W}_o \mathcal{\mathbf{a}}_{s,j}^t + \mathbf{U}_o (\mathbf{r}_{s,j}^t \odot \tilde{\mathbf{i}}_{s,j}^{t-1})),\\
\tilde{\mathbf{i}}_{s,j}^t &= (1-\mathbf{z}_{s,j}^t)\odot \tilde{\mathbf{i}}_{s,j}^{t-1} + \mathbf{z}_{s,j}^t \odot \hat{\mathbf{i}}_{s,j}^t,
\end{align}

where $\mathbf{A}^{in}_{s,j:}$, $\mathbf{A}^{out}_{s,j:} \in \mathbb{R}^{1\times l}$ depicts the $j$-th row of $\mathbf{A}^{in}_{s}$ and $\mathbf{A}^{out}_{s}$ respectively,  $\mathbf{H}_1$, $\mathbf{H}_2 \in \mathbb{R}^{d \times d}$, $\mathbf{W}_{(.)}$ and $\mathbf{U}_{(.)}$ are trainable parameters, $\sigma (.)$ is the sigmoid function, and $\odot$ is the element-wise multiplication operator.

To incorporate sequential information of item interactions, we optionally learn \textit{position embeddings} and add them to item embeddings to effectively obtain position-aware item (and subsequently session) embeddings.
The final embeddings for items in a session are computed as $\tilde{\mathbf{i}}^{\tau,p}_{s,j} = \tilde{\mathbf{i}}^\tau_{s,j} + \mathbf{p}_j$, where $\mathbf{p}_j \in \mathbb{R}^d$ is embedding vector for position $j$ obtained via a lookup over the position embeddings matrix $\mathbf{P} = [\mathbf{p}_1,\mathbf{p}_2,\ldots,\mathbf{p}_L]^T \in \mathbb{R}^{L\times d}$, where $L$ denotes the maximum length of any input session such that position $l\leq L$.

The soft-attention weight of the $j$-th item in session $s$ is computed as
$\alpha_j = \mathbf{q}^T\text{sigmoid}(\mathbf{W}_1\tilde{\mathbf{i}}^{\tau,p}_{s,l} + \mathbf{W}_2\tilde{\mathbf{i}}^{\tau,p}_{s,j} + \mathbf{c})$,
where $\mathbf{q}, \mathbf{c}\in \mathbb{R}^d$, $\mathbf{W}_1,\mathbf{W}_2 \in \mathbb{R}^{d\times d}$.
The $\alpha_j$s are further normalized using softmax.
An intermediate session embedding $\mathbf{s}'$ is computed as:
$\mathbf{s}' = \sum_{j=1}^{t}\alpha_j\tilde{\mathbf{i}}^{\tau,p}_{s,j}$.
The session embedding $\mathbf{s}$ is a linear transformation over the concatenation of intermediate session embedding $\mathbf{s}'$ and the embedding of most recent item $\tilde{\mathbf{i}}^{\tau,p}_{s,l}$, s.t. $\mathbf{s} = \mathbf{W}_3[\mathbf{s}';\tilde{\mathbf{i}}^{\tau,p}_{s,l}]
$, where $\mathbf{W}_3\in \mathbb{R}^{d\times2d}$.

The final recommendation scores for the $m$ items are computed as per Eq. \ref{eq:softmax}.
Note that while the session-graph embedding is obtained using item embeddings $\tilde{\mathbf{i}}^{\tau,p}_{s,t}$ that are aware of the session-graph and sequence, the normalized item embeddings $\tilde{\mathbf{i}}_j$ ($j=1\ldots m$) independent of a particular session are used to compute the recommendation scores.

\section{Experimental Evaluation}
\textbf{Dataset Details:}
We evaluate NISER on publicly available benchmark datasets: i) Yoochoose (YC), ii) Diginetica (DN), and iii) RetailRocket (RR).
The YC\footnote{http://2015.recsyschallenge.com/challege.html} dataset is from RecSys Challenge 2015, which contains a stream of user clicks on an e-commerce website within six months. 
Given the large number of sessions in YC, the recent 1/4 and 1/64 fractions of the training set are used to form two datasets: YC-1/4 and YC-1/64, respectively, as done in \cite{wu2018session}.
The DN\footnote{http://cikm2016.cs.iupui.edu/cikm-cup} dataset is transactional data from CIKM Cup 2016 challenge. 
The RR\footnote{https://www.dropbox.com/sh/dbzmtq4zhzbj5o9/AACldzQWbw-igKjcPTBI6ZPAa?dl=0} dataset is from an e-commerce personalization company retailrocket, which published dataset with six month of user browsing activities.
\\
\textbf{Offline and Online setting:} 
We consider two evaluation settings: i. offline and ii. online.
For evaluation in offline setting, we consider static splits of train and test as used in \cite{wu2018session} for YC and DN.
For RR, we consider sessions from last 14 days for testing and remaining 166 days for training. 
The statistics of datasets are summarized in Table \ref{tab:datastats}.
For evaluation in online setting, we re-train the models every day for 2 weeks (number of sessions per day for YC is much larger, so we evaluate for 1 week due to computational constraints) by appending the sessions from that day to the previous train set, and report the test results of the trained model on sessions from the subsequent day.\\
\textbf{NISER and its variants:}
We apply our approach over GNN and adapt code\footnote{https://github.com/CRIPAC-DIG/SR-GNN} from \cite{wu2018session} with suitable modification described later.
We found that for sessions with length $l>10$, considering only the most recently clicked 10 items to make recommendations worked consistently better across datasets. 
We refer to this variant as GNN+ and use this additional pre-processing step in all our experiments.
\begin{table}[t]
	\caption{Statistics of the datasets used for offline experiments. \label{tab:datastats}}
	\vspace{-2mm}
	\scalebox{0.9}{
		\begin{tabular}{|p{0.25\linewidth}|c|c|c|c|}
			\hline			
			\textbf{Statistics} & \textbf{DN} & \textbf{RR} & \textbf{YC-1/64} & \textbf{YC-1/4}\\
			\hline			
			\#train sessions&0.7 M&0.7 M&0.4 M&0.6 M\\
			\#test sessions&60,858&60,594&55,898&55,898\\
			\#items&43,097&48,759&16,766&29,618\\
			Average length&5.12&3.55&6.16&5.17\\
			\hline
		\end{tabular}}
		\vspace{-4mm}
	\end{table}
We consider enhancement over GNN+, and proposed following variants of the embedding normalization approach: 
\begin{itemize}
\item \textit{Normalized Item Representations (NIR)}: only item embeddings are normalized and scale factor $\sigma$ is not used\footnote{$\tilde{\mathbf{i}}_k^T\mathbf{s}$ is not restricted to [-1,1], in general $||\mathbf{s}||_{_2}\neq1$.},
\item \textit{Normalized Item and Session Representations (NISER)}: both item and session embeddings are normalized,
\item \textit{NISER+}: NISER with position embeddings and dropout applied to input item embeddings.
\end{itemize}

\textbf{Hyperparameter Setup:} 
Following \cite{wu2018session}, 
we use $d=100$ and learning rate of 0.001 with Adam optimizer. 
We use 10\% of train set which is closer to test test in time as hold-out validation set for hyperparameter tuning including scale factor $\sigma$.
We found $\sigma = 16.0$ to work best across most models trained on respective hold-out validation set chosen from \{$4.0,9.0,16.0,25.0$\}, and hence, we use the same value across datasets for consistency. 
We use dropout probability of 0.1 on dimensions of item embeddings in NISER+ across all models.
Since the validation set is closer to the test set in time, therefore, it is desirable to use it for training the models. 
After finding the best epoch 
via early stopping based on performance on validation set, we re-train the model for same number of epochs on combined train and validation set.
We train five models for the best hyperparameters with random initialization, and report average and standard deviation of various metrics for all datasets except for YC-1/4 where we train three models (as it is a large dataset and takes around ~15 hours for training one model). 
\\
\textbf{Evaluation Metrics:}
We use same evaluation metrics Recall@K and Mean Reciprocal Rank (MRR@K) as in \cite{wu2018session} with $K=20$.
\textbf{Recall@K} represents the proportion of test instances which has desired item in the top-K items.
\textbf{MRR@K} (Mean Reciprocal Rank) is the average of reciprocal ranks of desired item in recommendation list. 
Large value of MRR indicates that desired item is in the top of the recommendation list.
For evaluating popularity bias, we consider the following metrics as used in \cite{abdollahpouri2019managing}:
	\begin{figure*}[t]
			\subfigure[DN]{\includegraphics[trim={0.250cm 0cm 0.4cm 0cm},clip,width=0.22\textwidth]{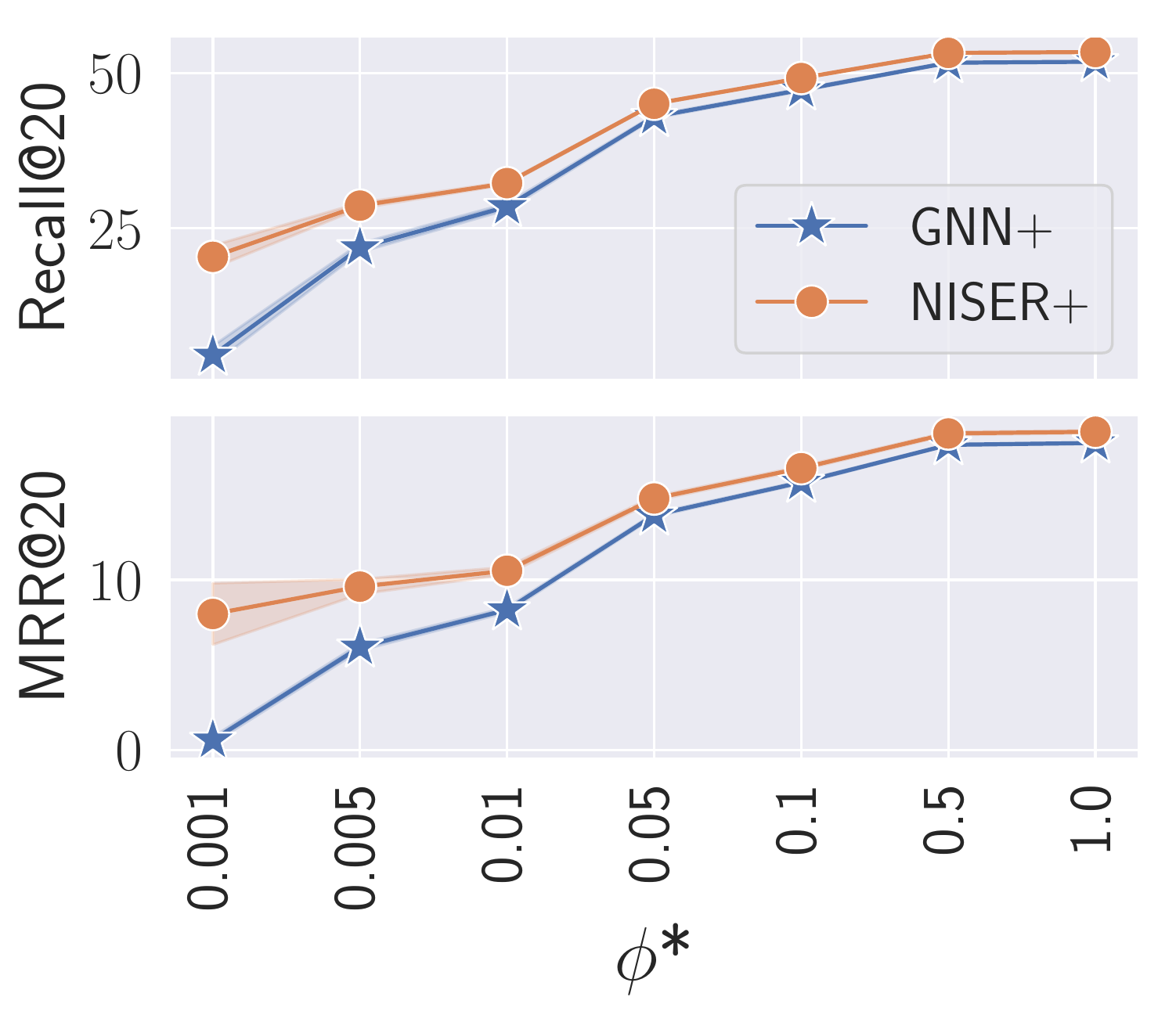}}
			\subfigure[RR]{\includegraphics[trim={0.250cm 0cm 0.4cm 0cm},clip,width=0.21\textwidth]{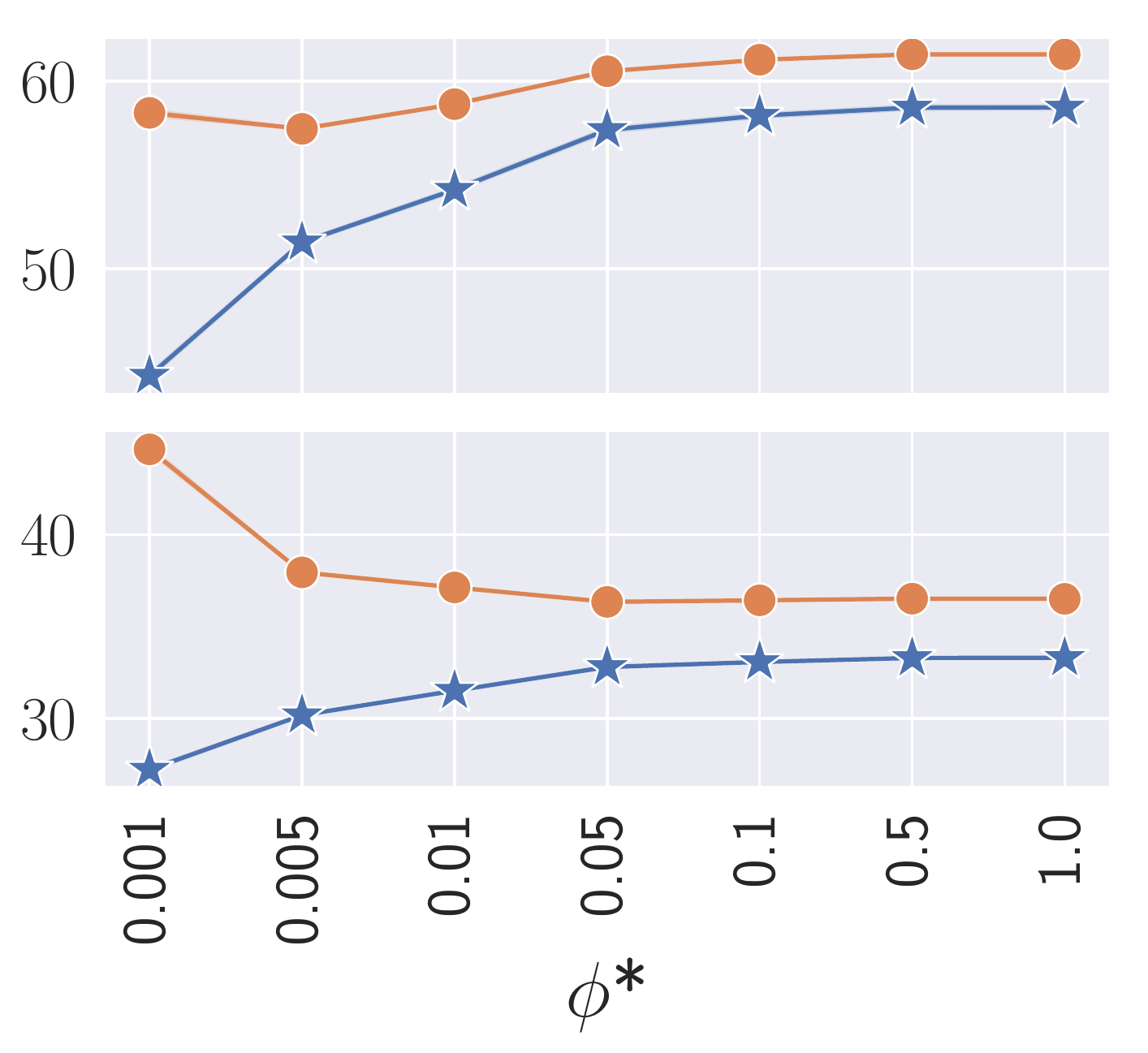}}
			\subfigure[YC-1/64]{\includegraphics[trim={0.250cm 0cm 0.4cm 0cm},clip,width=0.21\textwidth]{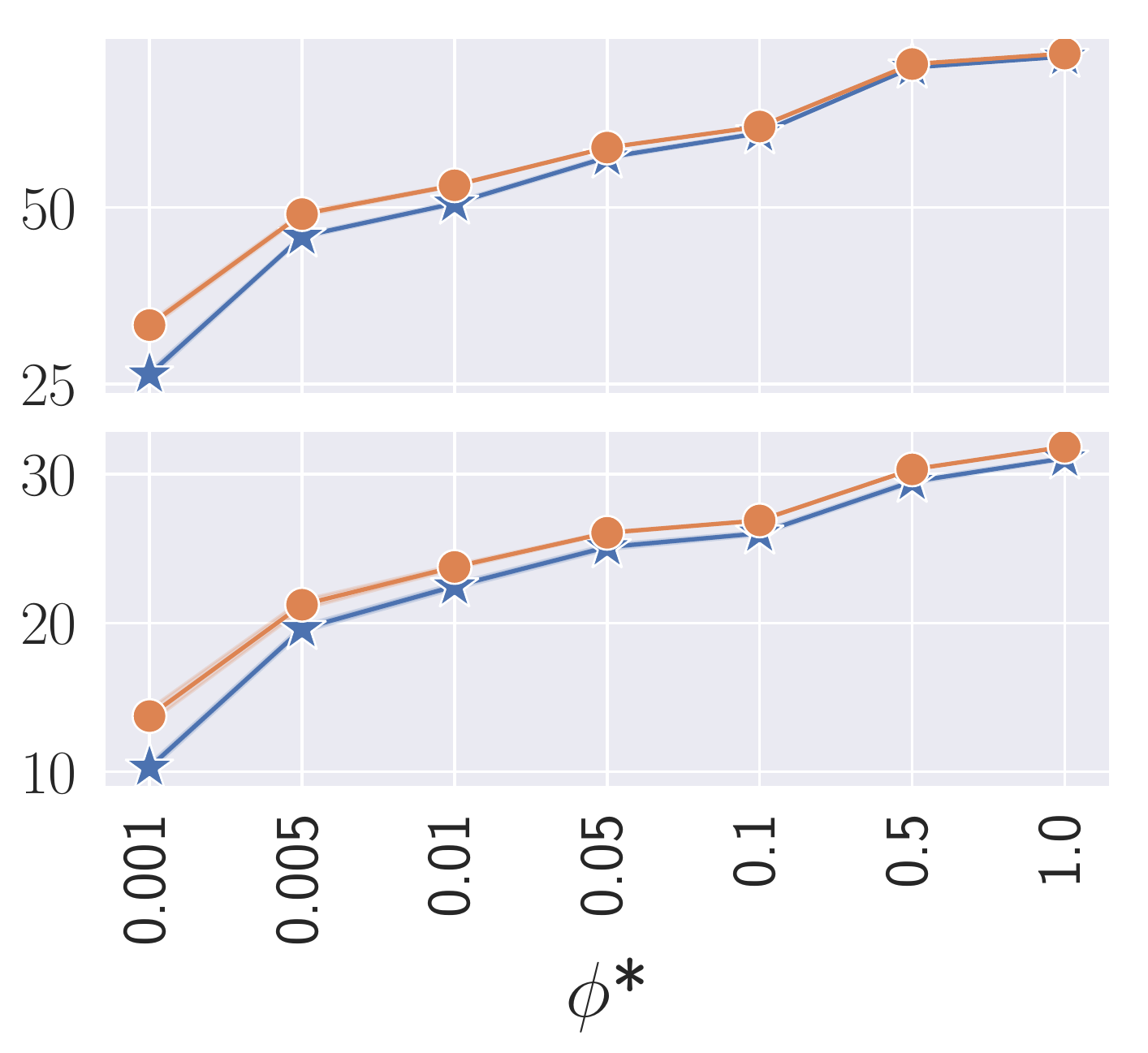}}
			\subfigure[YC-1/4]{\includegraphics[trim={0.250cm 0cm 0.4cm 0cm},clip,width=0.21\textwidth]{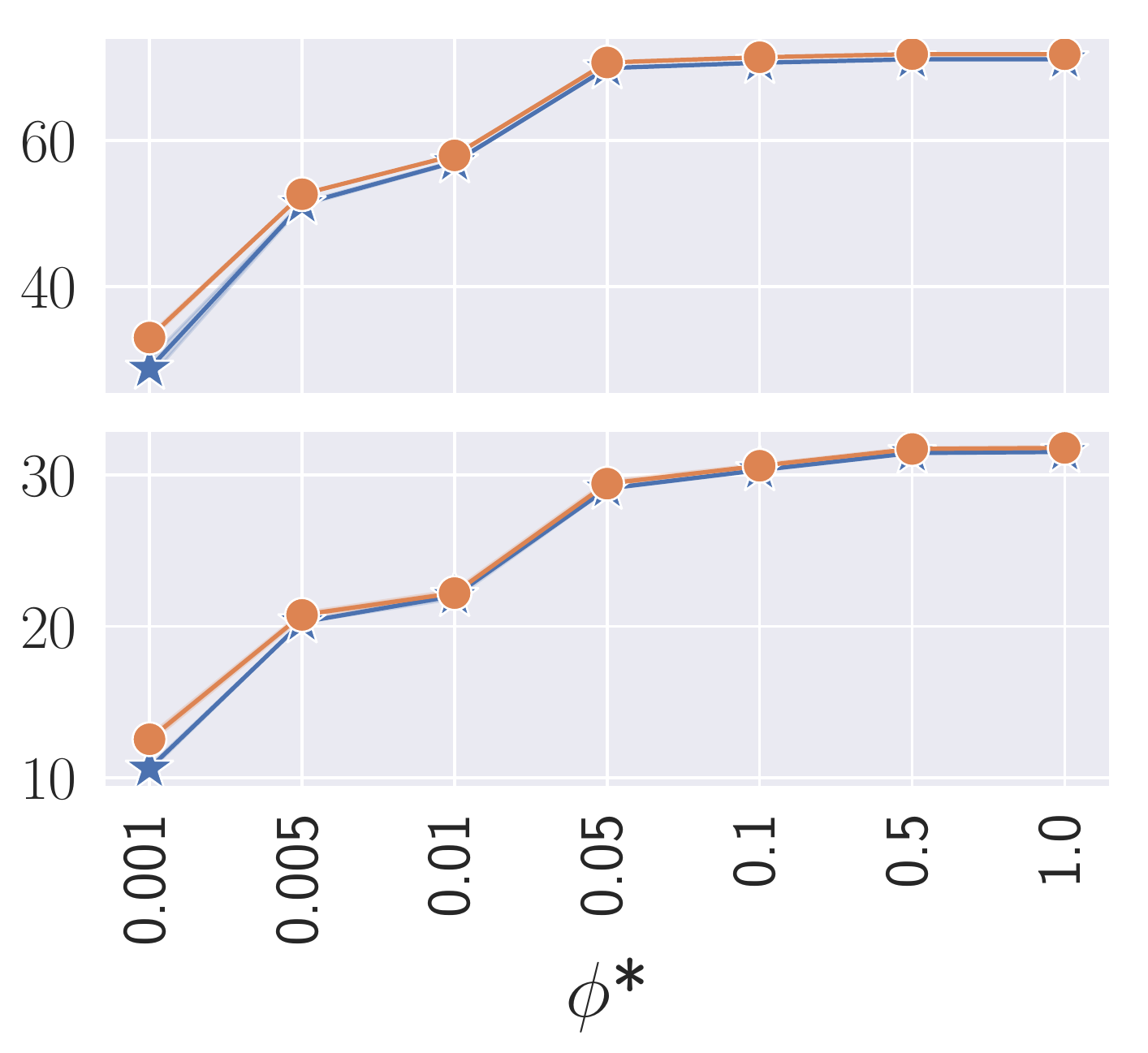}}
			\vspace{-2mm}
			\caption{Offline setting evaluation: Recall@20 and MRR@20 with varying $\phi^*$ indicating larger gains by using NISER+ over GNN+ for less popular items. \label{fig:recall-vs-phi}}
		\end{figure*}

		\begin{figure*}[t]
			\subfigure[DN]{\includegraphics[trim={0.0cm 0cm 0.0cm 0cm},clip,width=0.29\textwidth]{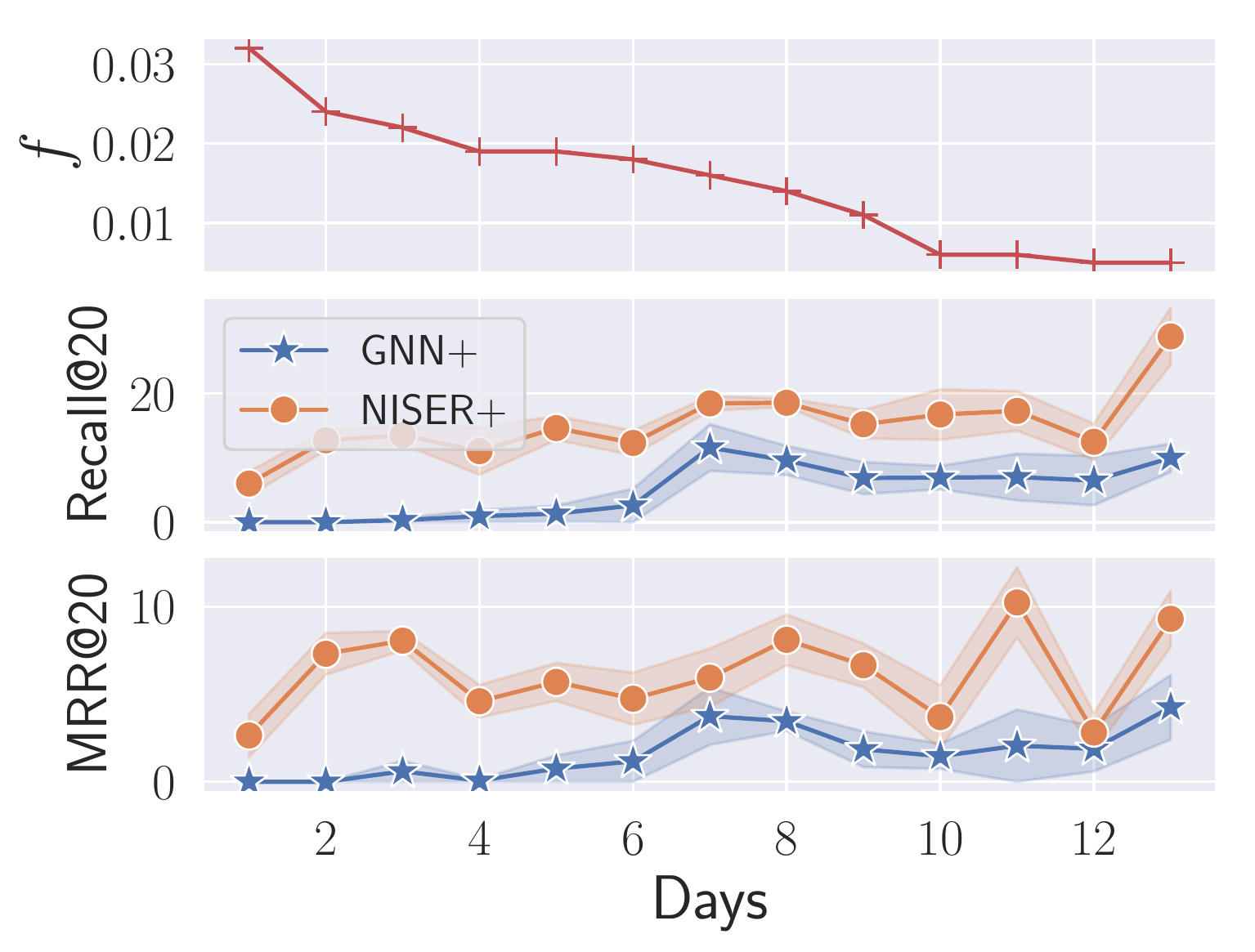}}
			\subfigure[RR]{\includegraphics[trim={0.0cm 0cm 0.0cm 0cm},clip,width=0.265\textwidth]{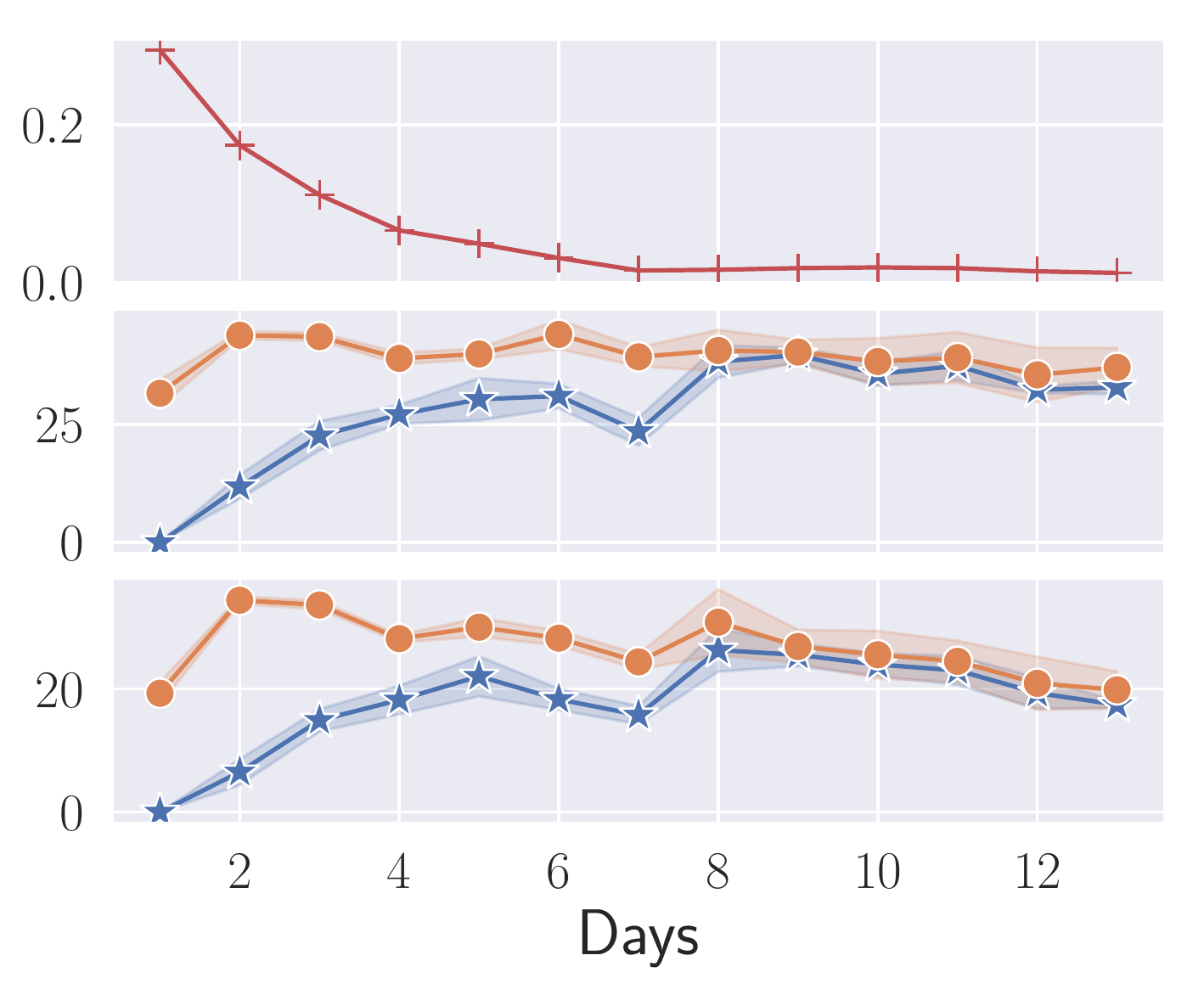}}
			\subfigure[YC]{\includegraphics[trim={0.0cm 0cm 0.0cm 0cm},clip,width=0.27\textwidth]{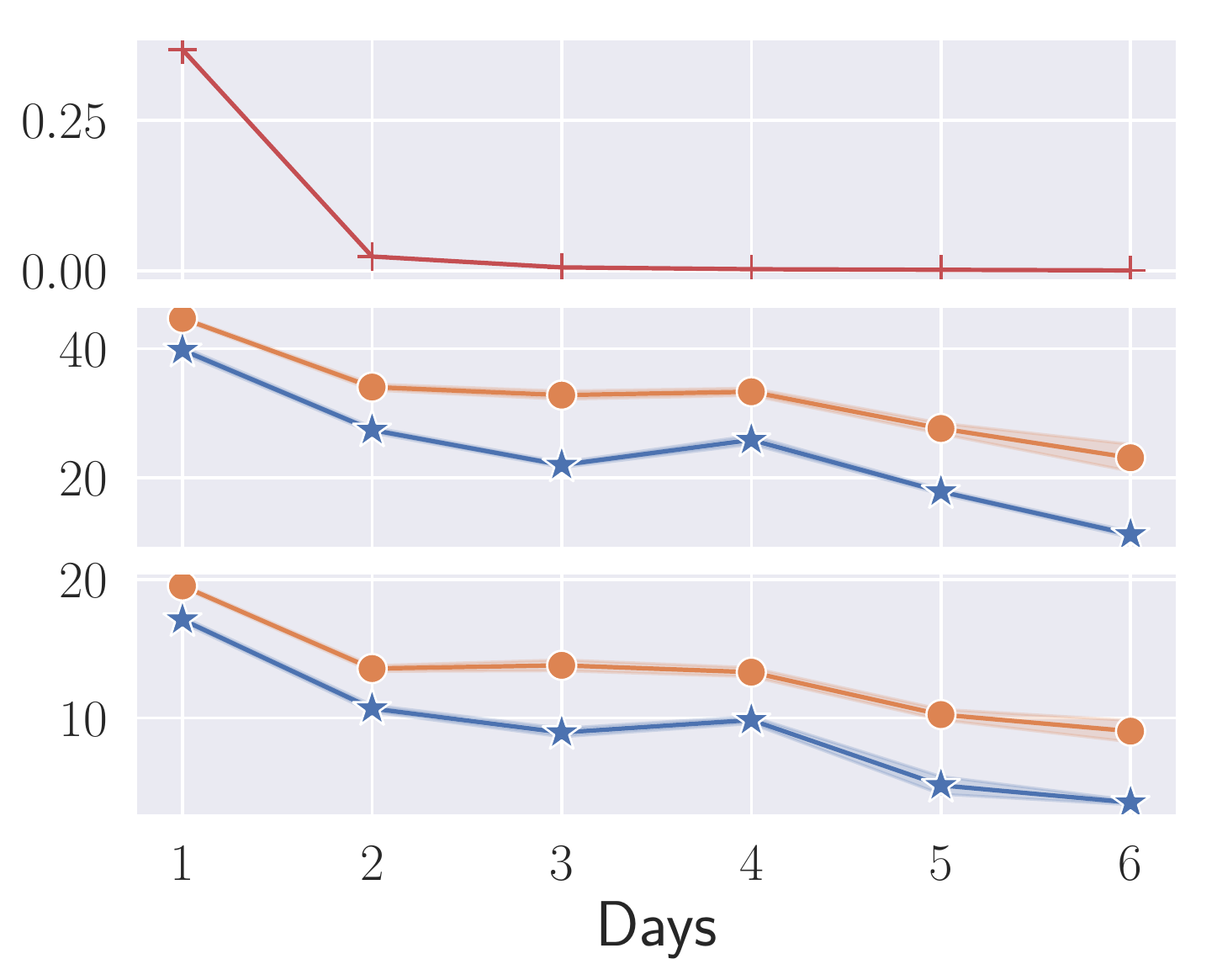}}
			\vspace{-2mm}
			\caption{Online setting evaluation: Recall@20 and MRR@20 for sessions where target item is one of the less popular newly introduced items from the previous day. $f$ denotes the fraction of such sessions in the training set for $\phi^*=0.01$. Standard deviations over five models are shown in lighter-shaded region around the solid lines. \label{fig:plot-online}}
		\end{figure*}
	
	\textbf{Average Recommendation Popularity (ARP)}: 
	This measure calculates the average popularity of the recommended items in each list given by:
	\begin{equation}\label{eq:arp}
	ARP=\frac{1}{|S|}\sum_{s \in S} \frac{\sum_{i \in L_s}\phi(i)}{K},
	\end{equation}
	where $\phi(i)$ is popularity of item $i$, i.e. the number of times item $i$ appears in the training set, $L_s$ is the recommended list of items for session $s$, and $|S|$ is the number of sessions in the test set.
	An item $i$ belongs to the set $\Gamma_{\phi^*}$ of long-tail items or less popular items if $\frac{\phi(i)}{max_i\phi(i)} \leq \phi^*$. 
	We evaluate the performance in terms of Recall@20 and MRR@20 for the sessions with target item in the set $\Gamma_{\phi^*}$ by varying $\phi^*$.
	\subsection{Results and Observations}

		\begin{table}[t]
		\caption{Offline setting evaluation: NISER+ versus GNN+ in terms of Average Recommendation Popularity (ARP). Lower values of ARP indicate lower popularity bias. \label{tab:pop-bias}}
		\scalebox{0.82}{			
			\begin{tabular}{|c|c|c|c|c|}
				\hline
				\textbf{Method}&\textbf{DN}&\textbf{RR}&\textbf{YC-1/64}&\textbf{YC-1/4}\\
				\hline
				GNN+ &495.25$\pm$2.52&453.39$\pm$8.97&4128.54$\pm$27.80&17898.10$\pm$126.93\\
				NISER+ &\textbf{487.31$\pm$0.30}&\textbf{398.53$\pm$3.09}&\textbf{3972.40$\pm$41.04}&\textbf{16683.52$\pm$120.74}\\
				\hline
			\end{tabular}}
			\vspace{-2mm}
		\end{table}
	
			\begin{table}[t]
				
				{\caption{NISER+ versus other benchmark methods in offline setting. Numbers after $\pm$ are standard deviation values over five models. \label{tab:gnn_all_r}}}
				\scalebox{0.85}{
					
					\begin{tabular}{|c|c|c|c|c|}
						\hline
						\textbf{Method}&\textbf{DN}&\textbf{RR}&\textbf{YC-1/64}&\textbf{YC-1/4}\\
						\hline
						\multicolumn{5}{|c|}{\textbf{Recall@20}}\\
						\hline
						SKNN \cite{jannach2017recurrent}&48.06&56.42&63.77&62.13\\
						STAN \cite{garg2019sequence} &50.97&59.80&69.45&70.07\\
						\hline
						GRU4REC \cite{hidasi2015session}&29.45&-&60.64&59.53\\
						NARM \cite{li2017neural}&49.70&-&68.32&69.73\\
						STAMP \cite{liu2018stamp}&45.64&53.94&68.74&70.44\\
						GNN \cite{wu2018session}&51.39$\pm$0.38&57.63$\pm$0.15&70.54$\pm$0.14&70.95$\pm$0.04\\
						GNN+&51.81$\pm$0.11&58.59$\pm$0.10&70.85$\pm$0.08&71.10$\pm$0.07\\
						\hline			
						NIR&52.40$\pm$0.06&60.67$\pm$0.08&71.12$\pm$0.05&71.32$\pm$0.11\\
						NISER&52.63$\pm$0.09&60.85$\pm$0.06&70.86$\pm$0.15&71.69$\pm$0.03\\
						NISER+&\textbf{53.39$\pm$0.06}&\textbf{61.41$\pm$0.09}&\textbf{71.27$\pm$0.05}&\textbf{71.80$\pm$0.09}\\			
						\hline\hline
						\multicolumn{5}{|c|}{\textbf{MRR@20}}\\
						\hline
						SKNN \cite{jannach2017recurrent}&16.95&33.16&25.22&24.82\\
						STAN \cite{garg2019sequence}&18.48&35.32&28.74&28.89\\
						\hline
						GRU4REC \cite{hidasi2015session}&8.33&-&22.89&22.60\\
						NARM \cite{li2017neural}&16.17&-&28.63&29.23\\
						STAMP \cite{liu2018stamp}&14.32&28.49&29.67&30.00\\
						GNN \cite{wu2018session}&17.79$\pm$0.16&32.74$\pm$0.09&30.80$\pm$0.09&31.37$\pm$0.13\\
						GNN+&18.03$\pm$0.05&33.29$\pm$0.03&30.84$\pm$0.10&31.51$\pm$0.05\\
						\hline
						NIR &18.52$\pm$0.06&35.57$\pm$0.05&30.99$\pm$0.10&31.73$\pm$0.11\\
						NISER &18.27$\pm$0.10&36.09$\pm$0.03&31.50$\pm$0.11&\textbf{31.80$\pm$0.12}\\			
						NISER+&\textbf{18.72$\pm$0.06}&\textbf{36.50$\pm$0.05}&\textbf{31.61$\pm$0.02}&31.77$\pm$0.10\\
						\hline
					\end{tabular}}
					\vspace{-2mm}
				\end{table}
				
				\begin{table}
					
					\caption{Ablation results for NISER+ indicating that normalization of embeddings (L$_2$ norm) contributes the most to performance improvement. Here PE: Position Embeddings. \label{tab:gnn_plus_ablation_recall}}
					\scalebox{0.85}{
						\begin{tabular}{|p{0.2\linewidth}|c|c|c|c|}
							\hline
							
							\textbf{Method}&\textbf{DN}&\textbf{RR}&\textbf{YC-1/64}&\textbf{YC-1/4}\\
							\hline			
							\multicolumn{5}{|c|}{\textbf{Recall@20}}\\
							\hline
							NISER+&\textbf{53.39$\pm$0.06}&\textbf{61.41$\pm$0.09}&\textbf{71.27$\pm$0.05}&71.80$\pm$0.09\\			
							-L$_2$ norm &52.23$\pm$0.10&59.16$\pm$0.10&71.10$\pm$0.09&71.46$\pm$0.19\\
							-Dropout&52.81$\pm$0.12&60.99$\pm$0.09&71.07$\pm$0.13&\textbf{71.90$\pm$0.03}\\
							-PE&53.11$\pm0.12$&61.22$\pm$0.03&71.13$\pm$0.04&71.70$\pm$0.11\\
							\hline\hline
							\multicolumn{5}{|c|}{\textbf{MRR@20}}\\
							\hline
							NISER+&\textbf{18.72$\pm$0.06}&\textbf{36.50$\pm$0.05}&31.61$\pm$0.02&31.77$\pm$0.10\\
							-L$_2$ norm&18.11$\pm$0.05&33.78$\pm$0.04&30.90$\pm$0.07&31.49$\pm$0.07\\
							-Dropout&18.43$\pm$0.11&35.99$\pm$0.02&31.56$\pm$0.06&\textbf{31.93$\pm$0.17}\\
							-PE &18.60$\pm$0.09&36.32$\pm$0.03&\textbf{31.68$\pm$0.05}&31.71$\pm$0.06\\
							\hline
						\end{tabular}}
						\vspace{-2mm}
					\end{table}
					
					\textbf{(1)} \textbf{NISER+ reduces popularity bias in GNN+ in :} 
					Table \ref{tab:pop-bias} shows that ARP for NISER+ is significantly lower than GNN+ indicating that NISER+ is able to recommend less popular items more often than GNN+, thus reducing popularity bias.
					Furthermore, from Fig. \ref{fig:recall-vs-phi}, we observe that NISER+ outperforms GNN+ for sessions with less popular items as targets (i.e. when $\phi^*$ is small), with gains 13\%, 8\%, 5\%, and 2\% for DN, RR, YC-1/64, and YC-1/4 respectively for $\phi^*=0.01$ in terms of Recall@20. 
					Similarly, gains are 28\%, 18\%, 6\%, and 2\% in terms of MRR@20.
					Gains for DN and RR are high as compared to YC. This is due to the high value of $max_i\phi(i)$. If instead we consider $\phi^*=0.001$, gains are as high as 26\% and 9\% for YC-1/64 and YC-1/4 respectively in terms of Recall@20. Similarly, gains are as high as 34\% and 19\% in terms of MRR@20.
					We also note that NISER+ is at least as good as GNN+ even for the sessions with more popular items as targets (i.e. when $\phi^*$ is large).

					\noindent \textbf{(2)} \textbf{NISER+ improves upon GNN+ in online setting}
					for newly introduced items in the set of long-tail items $\Gamma_{\phi^*}$. These items have small number of sessions available for training at the end of the day they are launched.
					From Fig. \ref{fig:plot-online}, we observe that for the less popular newly introduced items, NISER+ outperforms GNN+ for sessions where these items are target items on the subsequent day. 
					This proves the ability of NISER+ to recommend new items on very next day, due to its ability to reduce popularity bias. 
					Furthermore, for DN and RR, we observe that during initial days, when training data is less, GNN+ performs poorly while performance of NISER+ is relatively consistent across days indicating potential regularization effect of NISER on GNN models in less data scenarios. 
					Also, as days pass by and more data is available for training, performance of GNN+ improves with time but still stays significantly below NISER+.
					For YC, as days pass by, the popularity bias becomes more and more severe (as depicted by very small value for $f$, i.e. the fraction of sessions with less popular newly introduced items) such that the performance of both GNN+ and NISER+ degrades with time. However, importantly, NISER+ still performs consistently better than GNN+ on any given day as it better handles the increasing popularity bias.
					
					\noindent \textbf{(3)} \textbf{NISER and NISER+ outperform GNN and GNN+ in offline setting:}
					From Table \ref{tab:gnn_all_r}, we observe that NISER+ shows consistent and significant improvement over GNN in terms of Recall and MRR, establishing a \textit{new state-of-the-art} in SR. 
					
					We also conduct an \textbf{ablation study} (removing one feature of NISER+ at a time) to understand the effect of each of the following features of NISER+: i. L$_2$ norm of embeddings, ii. including position embeddings, and iii. applying dropout on item embeddings. As shown in Table \ref{tab:gnn_plus_ablation_recall}, we observe that L$_2$ norm is the most important factor across datasets while dropout and position embeddings contribute in varying degrees to the overall performance of NISER+.
					
\section{Discussion}
In this work, we highlighted that the typical item-frequency distribution with long tail leads to popularity bias in state-of-the-art deep learning models such as GNNs \cite{wu2018session} for session-based recommendation. 
We then argued that this is partially related to the `radial' property of the softmax loss that, in our setting, implies that the norm for popular items will likely be larger than the norm of less popular items.
We showed that learning the representations for items and session-graphs by optimizing for cosine similarity instead of inner product can help mitigate this issue to a large extent.
Importantly, this ability to reduce popularity bias is found to be useful in the online setting where the newly introduced items tend to be less popular and are poorly modeled by existing approaches.
We observed significant improvements in overall recommendation performance by normalizing the item and session-graph representations and improve upon the existing state-of-the-art results.
In future, it would be worth exploring NISER to improve other algorithms like STAMP \cite{liu2018stamp} that rely on similarity between embeddings for items, sessions, users, etc.

\bibliographystyle{ACM-Reference-Format}
\bibliography{gnn-plus,sigir}


\end{document}